\documentclass[nofootinbib,letter, scriptaddress, twocolumn, prd, showpacs,showkeys]{revtex4-1}
	\usepackage{amsmath}
        \usepackage{mathrsfs}
	\usepackage{amssymb}
	\usepackage{graphicx} 
	\usepackage{makeidx}
	\usepackage{amsfonts}
	\usepackage[ansinew]{inputenc}
	\usepackage[usenames, dvipsnames]{pstricks}
	\usepackage{epsfig}
	\usepackage{epstopdf}
	\usepackage{pst-grad} 
	\usepackage{pst-plot} 
	\usepackage[colorlinks, hyperindex]{hyperref}
	\colorlet{Mycolor1}{yellow!10!black!90!}
	\hypersetup
	{
		colorlinks, %
		citecolor= Red, %
		linkcolor=blue, %
		urlcolor=WildStrawberry, %
	}

\makeindex
\begin{document}

\title{{\Large\bf Equivalence of Einstein and Jordan frames in quantized cosmological models}}
\author{Sachin Pandey}
\email{sp13ip016@iiserkol.ac.in}
\affiliation{Department of Physical Sciences, \\
Indian Institute of Science Education and Research Kolkata, \\
Mohanpur, West Bengal 741246, India.}
\author{Sridip Pal} 
\email{sridippaliiser@gmail.com; srpal@ucsd.edu}
\affiliation{Department of Physics;
University of California, San Diego\\
9500 Gilman Drive, La Jolla, CA 92093, USA}
\author{Narayan Banerjee}
\email{narayan@iiserkol.ac.in}
\affiliation{Department of Physical Sciences, \\
Indian Institute of Science Education and Research Kolkata, \\
Mohanpur, West Bengal 741246, India.}
\begin{abstract}
\begin{center}
{\bf Abstract}
\end{center}
The present work shows that the mathematical equivalence of Jordan frame and its conformally transformed version, the Einstein frame, so far as Brans-Dicke theory is concerned, survives a quantization of cosmological models in the theory. We work with the Wheeler-deWitt quantization scheme and take up quite a few anisotropic cosmological models as examples. We effectively show that the transformation from Jordan to Einstein frame is a canonical one and hence two frames are equivalent description of same physical scenario.
\end{abstract}
\maketitle

\section{Introduction}
Brans-Dicke theory (BDT) is one of the modified theories of gravity where a scalar field is non-minimally coupled to gravity in the form of a product term of $\phi$ and the Ricci scalar $R$. The presence of non minimal coupling in Einstein-Hilbert action signals departure from its closet cousin, General Relativity (GR). In its defining frame\cite{brans}, the so called Jordan frame, the non-minimal coupling is manifest. Now one can effect a clever conformal transformation to sweep the non-minimal coupling under the rug\cite{dicke}, i.e in this conformally transformed frame, so called Einstein frame, the geometry is only minimally coupled to the scalar field and action looks like that in GR. The price of recovering minimal coupling is that mass and anything with  mass dimension transforms while going to Einstein frame, hence the geodesic equations, which is valid in Jordan frame is no longer valid in Einstein frame. Thus, the physical content of BDT is quite different in two frames and one has to declare one of the frame to be the physical one while the other is not. Chiba and Yamaguchi\cite{chiba} showed that cosmological parameters in these frames are indeed different and hence a matter of concern. Faraoni and Gunzig\cite{faraoni} argued in favor of Jordan frame in the classical regime, in the context of gravitational waves.  On the other hand, Cho argued that the Einstein frame is more trustworthy for a physical description of gravity\cite{cho}.  \\

However, classically one is allowed to do the calculations in either of the frames, but to have physical insight to the theory, one must transform it back to the physically relevant frame. The inverse transformation needs to be well behaved for this. The boost comes from the investigations by Salgado who resolved the mismatch between the Cauchy problem in the two frames\cite{salgado}. Faraoni and Nadeau also showed that at a classical level, the alleged nonequivalence of the two frames is actually a matter of interpretation\cite{faraoni2}.\\

Although the physical inequivalence and henceforth the question of which frame being the physical one is a moot point for decades\cite{morgan, magnano, sokol, faraoni3, cho2, ippo}, mathematical equivalence of these two frames at a classical level has almost been taken for granted. By mathematical equivalence, we mean following: we evaluate a quantity $T_{J}$ in Jordan frame, evaluate the same quantity in Einstein frame and obtain $T_{E}$,  mathematical equivalence is defined to be the statement that $T_{E}$ is just a transformed version of $T_{J}$. Given this scenario, it is meaningful to explore whether this mathematical equivalence survives a quantization process. \\

In the light of recent resurgence in Wheeler-deWitt (WdW) quantization \cite{sp1,sp2,sp3,sp4,sp5,sachin, alvarenganew} scheme, we have a natural framework to answer the question of equivalence at the quantum level. In fact, within WdW quantization scheme, there have been claims of non-equivalence at quantum level in literature \cite{nb, abhik, odint, ma}. Very recently Kamenshchik and Steinwachs\cite{kamen}, with an estimate of the one loop divergence in the two frames, showed that the frames are not equivalent. Nonetheless, the debate seems to be open enough. The hint of quantum equivalence is also speculated in \cite{sp6} and further claimed to be true for isotropic homogeneous model with scale invariant matter content in \cite{almida}, generically in \cite{pucheu}. \\

The purpose of this work is to settle the issue and pinpoint the origin of non-equivalence, which turns out to be an inconsistent operator ordering in the two frames. To be specific, we explicitly choose a consistent set of parametrization and operator ordering to show equivalence for homogeneous models without matter content in the present work. Hence, the apparent mismatch reported in literature is not really due to some ``physical"  quantum effects, rather an  inconsistent operator ordering is responsible for the illusion.\\

The rest of the work is organized as follows. In section II, we analyze the model with zero spatial curvature: Bianchi I. The section III deals with models having constant nonzero spatial curvature i.e Bianchi-V and Bianchi-IX. The model with time dependent spatial curvature is dealt with in section IV. We conclude our work with a general discussion and a sketch of generic way to prove equivalence for all homogeneous models. It deserves mention that our result agrees with the promised, yet to be published paper \cite{romero1}, mentioned in \cite{pucheu}.

\section{Model with zero spatial curvature: Bianchi I}
\subsection{Jordan frame}
In the absence of any matter field, action for Brans Dicke theory in Jordan frame is given by
\begin{equation}\label{actionb1}
A_{J}=\int d^{4}x\mathcal{L}=\int d^4 x \sqrt{-g} \bigg{[} \phi R + \frac{\omega}{\phi}\partial_\mu \phi \partial^\nu \phi \bigg{]} .
\end{equation}

The Bianchi-I metric is described by 
\begin{equation}
\label{b1metric}
ds^2 = n^2dt^2 - a^2(t)dx^2-b^2(t)dy^2-c^2(t)dz^2
\end{equation}
where $a,b,c$ are the scale factors along three spatial direction. They encode the anisotropy present in the metric. \\

The Lagrangian arising out of Eq.~\eqref{actionb1} can be explicitly written as: 
\begin{equation}
L_{J}=-\frac{2\phi abc}{n}\left[ \frac{\dot{a}}{a} \frac{\dot{b}}{b}+ \frac{\dot{a}}{a} \frac{\dot{c}}{c}+\frac{\dot{b}}{b} \frac{\dot{c}}{c}+\frac{\dot{a}}{a}\frac{\dot{\phi}}{\phi}+ \frac{\dot{b}}{b} \frac{\dot{\phi}}{\phi}+\frac{ \dot{c}}{c} \frac{\dot{\phi}}{\phi}  - \frac{\omega}{2}\frac{\dot{\phi}^2}{\phi^{2}}\right],
\end{equation}
where an overhead dot signifies a differentiation with respect to $t$. We reparametrize the scale factors as,
\begin{eqnarray}
\label{b1trans}
a(t)&=&e^{\sigma_0 +\sigma_+ +\sqrt{3}\sigma_-},\\
b(t)&=&e^{\sigma_0 +\sigma_+ -\sqrt{3}\sigma_-},\\
c(t)&=&e^{\sigma_0 -2\sigma_+},\\
\phi(t)&=&e^{\alpha},
\end{eqnarray}

which recast the Lagrangian in following form:
\begin{equation}\label{L1}
L_{J}=\frac{e^{\alpha+3\sigma_{0}}}{n}\left[-6(\dot{\sigma}_{0}^{2}-\dot{\sigma}_{+}^{2}-\dot{\sigma}_{-}^{2})-6\dot{\alpha}\dot{\sigma}_{0}+\omega\dot{\alpha}^{2}\right].
\end{equation}

Now to kill the cross-term $\dot{\alpha}\dot{\sigma}_{0}$, we do a change of variables as, (which is in fact a canonical one) \footnote{\textit{We thank Ott Vilson for point out to us that this is a canonical transformation even at quantum level.}}
\begin{eqnarray}
\label{conformal}\beta_{0}&=&\sigma_{0}+\frac{\alpha}{2},\\
\beta_{\pm}&=&\sigma_{\pm}.
\end{eqnarray}

which yield following Lagrangian
\begin{equation}
L_{J} = \frac{e^{-\frac{\alpha}{2}+ 3\beta_{0}}}{n}\left[-6\dot{\beta}_{0}^{2}+6\dot{\beta}_{+}^{2}+6\dot{\beta}_{-}^{2}+\frac{2\omega +3}{2}\dot{\alpha}^2\right],
\end{equation}
 and the corresponding Hamiltonian is given by 
\begin{equation}
H_{J} =  \frac{ne^{\frac{\alpha}{2} - 3 \beta_{0}}}{4}\left[-\frac{1}{6}(p_0^2-p_+^2-p_-^2)+\frac{2}{2\omega+3}p_\alpha^2\right],
\end{equation}
where $p_{0}$ and $p_{\pm}$ are momenta conjugate to $\beta_{0}$ and $\beta_{\pm}$ respectively while $p_{\alpha}$ is momentum conjugate to $\alpha$. \\

The transformation \eqref{conformal} looks like the one which takes us from Jordan to Einstein frame. But at this stage, they are to be thought merely as a canonical transformation. Hence, the Lagrangian and Hamiltonian above describes Brans-Dicke theory in a canonically equivalent frame of Jordan frame.

Upon variation with respect to $n$, we obtain Hamiltonian constraint: 
\begin{equation}\label{Ham}
\mathcal{H}=\left[-\frac{1}{6}(p_0^2-p_+^2-p_-^2)+\frac{2}{2\omega+3}p_\alpha^2\right]=0.
\end{equation}

If we now consider a canonical transformation $(\alpha,p_\alpha)$ to $(T,p_T)$ as
\begin{eqnarray}
T&=&\frac{\alpha}{p_\alpha},\\
p_T&=&\frac{p_{\alpha}^{2}}{2},
\end{eqnarray}
 then Eq. ~\eqref{Ham} can be rewritten as 
\begin{equation}
\mathcal{H}=-\frac{1}{6}\left(p_0^2-p_+^2-p_-^2-\frac{24}{2\omega+3}p_T\right).
\end{equation}

In the absence of a properly oriented scalar time parameter in the theory, we have used the evolution of the scalar field as the relevant time parameter. The method is the same as that suggested by Vakili\cite{vakili}. That $T$ has the proper orientation can be ascertained from its dependence on the cosmic time in the right direction. For a summary of the method, we refer to \cite{sp1}. \\

Upon quantization, we obtain Wheeler deWitt equation as follows,
\begin{equation}\label{master equation}
\left[\frac{\partial^2}{\partial \beta_0^2}-\frac{\partial^2}{\partial \beta_+^2}-\frac{\partial^2}{\partial \beta_-^2} - i\frac{24}{2\omega+3} \frac{\partial}{\partial T}\right]\psi=0.
\end{equation}

\subsection{Einstein Frame}
In Einstein frame, the transformed metric components are ${\bar{g}}_{\mu\nu} = \phi g_{\mu\nu}$, the action is given by
\begin{equation}
A_E=\int d^4x \sqrt{-\bar{g}}\left[\bar{R}+\frac{2\omega+3}{2}\partial_\mu \xi \partial^\nu \xi \right],
\end{equation}
where $\xi = ln\phi$. 
Again, Bianchi I metric is given by, 
\begin{equation}
ds^2 = \bar{n}^2(t)dt^2 - \bar{a}^2(t)dx^2-\bar{b}^2(t)dy^2-\bar{c}^2(t)dz^2.
\end{equation}
The Lagrangian can be written as 
\begin{equation}
L_E=\frac{e^{3r_0}}{\bar{n}}\left[-6(\dot{r}_{0}^2-\dot{r}_{+}^{2}-\dot{r}_{-}^2)+\frac{2\omega+3}{2}\dot{\xi}^2\right],
\end{equation}
where
\begin{eqnarray}
\bar{a}(t)&=&e^{r_0 +r_+ +\sqrt{3}r_-},\\
\bar{b}(t)&=&e^{r_0 +r_+ -\sqrt{3}r_-},\\
\bar{c}(t)&=&e^{r_0 -2r_{+}}
\end{eqnarray}
and we obtain the following Hamiltonian 
\begin{equation}
H_E = \frac{\bar{n}e^{ - 3 r_0}}{4}\left[-\frac{1}{6}(\bar{p}_0^2-\bar{p}_+^2-\bar{p}_-^2)+\frac{2}{2\omega+3}p_\xi^2\right].
\end{equation}
Now varying the action with respect to $\bar{n}$, one obtains the Hamiltonian constraint $\mathcal{H_E}=\frac{e^{  3 r_0}}{\bar{n}}H_E =0$. Again with similar canonical transformation as
\begin{eqnarray}
\bar{T}&=&\frac{\xi}{p_\xi},\\
p_{\bar{T}}&=&\frac{p_\xi^2}{2},
\end{eqnarray}
the WDW equation can be written as
\begin{equation}
\left[\frac{\partial^2}{\partial r_0^2}-\frac{\partial^2}{\partial r_+^2}-\frac{\partial^2}{\partial r_-^2} -i\frac{24}{2\omega+3} \frac{\partial}{\partial \bar{T}}\right]\bar{\psi}=0.
\end{equation}

One can easily show that $r_i=\beta_i$ and $T=\bar{T}$ from canonical transformations done in both the frame work. Hence it is quite obvious to see that WdW equation in both the frame is exactly same, thus, will lead to the similar wave packet if we revert back to following variables $\bar{a} =\sqrt \phi a,\bar{b} =\sqrt \phi b ,\bar{c}= \sqrt \phi c$. \\

If in Eq.~\eqref{b1metric} we put $a=b=c$, then in Eq.~\eqref{b1trans}, both $\sigma_+$ and $\sigma_-$ become zero and we recover the same result for a spatially flat isotropic cosmological model. 

\section{Models with constant nonzero spatial curvature}
Bianchi V and Bianchi IX models have constant spatial curvature, with a negative and a positive signature respectively.
\subsection{Bianchi V}
\subsubsection{Jordan frame}

The Bianchi-V metric is given by
\begin{equation}
ds^2 = n^2(t)dt^2 - a^2(t)dx^2-e^{2 m x}[b^2(t)dy^2+c^2(t)dz^2],
\end{equation}

where $m$ is a constant. We parametrize the scale factors in following manner,

\begin{eqnarray}
a(t)&=&e^{\sigma_0 },\\
b(t)&=&e^{\sigma_0 +\sqrt{3}(\beta_+ -\beta_-)},\\
c(t)&=&e^{\sigma_0 -\sqrt{3}(\beta_+ -\beta_-)},\\
\phi(t)& =& e^{\alpha}
\end{eqnarray}
 and write down the Lagrangian, which will have some cross-term like $\dot{\alpha}\dot{\sigma_{0}}$. Now to kill the cross-term, we define $\beta_{0}=\sigma_{0}+\frac{\alpha}{2}$, which is again a canonical transformation. This immediately recasts the Lagrangian in following form, 
\begin{widetext}
\begin{equation}
L_{J} = \frac{e^{-\frac{\alpha}{2} + 3 \beta_0}}{n}\left[-6\dot{\beta}_{0}^2+6(\dot{\beta}_{+}-\dot{\beta}_{-})^2-6e^{-2\beta_0}n^2m^2+\frac{2\omega +3}{2}\dot{\alpha}^2 \right].
\end{equation}
\end{widetext}
The nontrivial part of this parametrization is that the canonical transformation required to make the Lagrangian in a diagonal form is essentially same as conformal transformation to Einstein frame. This essentially means that conformal transformation is indeed canonical even at quantum level. There is no a priori reason for them to be same.  \\

The corresponding Hamiltonian can be given as 
\begin{equation}
H_{J} =  \frac{ne^{\alpha / 2 - 3 \beta_0}}{24}\left[-(p_0^2-p_+^2-144m^2e^{4\beta_0})+\frac{12}{2\omega+3}p_\alpha^2\right].
\end{equation}
Variation with respect to $n$ yields $\mathcal{H}=\frac{e^{-\alpha / 2 + 3 \beta}H}{n}=0$.\\

Now we consider a canonical transformation as before $(\alpha,p_\alpha)$ to $(T,p_T)$ given by following:
\begin{eqnarray}
T&=&\frac{\alpha}{p_\alpha},\\
p_T&=&\frac{p_\alpha^2}{2},
\end{eqnarray}
 so that Wheeler de Witt equation becomes
\begin{equation}
\left[\frac{\partial^2}{\partial \beta_0^2}-\frac{\partial^2}{\partial \beta_+^2}-144m^2e^{4\beta_0} -i\frac{24}{2\omega+3} \frac{\partial}{\partial T}\right]\psi=0.
\end{equation}
\subsubsection{Einstein Frame}
Again, in Einstein frame the action is given by
\begin{equation}
A_E=\int d^4x \sqrt{-\bar{g}}\left[\bar{R}+\frac{2\omega+3}{2}\partial_\mu \xi \partial^\nu \xi \right],
\end{equation}
where $\xi = ln\phi$ and the  Bianchi V metric is given by
\begin{equation}
ds^2 = \bar{n}^2(t)dt^2 - \bar{a}^2(t)dx^2-e^{2mx}[\bar{b}^2(t)dy^2+\bar{c}^2(t)dz^2].
\end{equation}
Now the Lagrangian can be written as 
\begin{equation}
L_E=\frac{e^{3r_0}}{\bar{n}}\left[-6\dot{r}_{0}^2+6(\dot{r}_{+}-\dot{r}_{-})^2-6e^{-2r_0}n^2m^2+\frac{2\omega+3}{2}\dot{\xi}^2\right],
\end{equation}
where
\begin{eqnarray}
\bar{a}(t)&=&e^{r_0 },\\
\bar{b}(t)&=&e^{r_0 +\sqrt{3}(r_+ - r_-)},\\
\bar{c}(t)&=&e^{r_0 -\sqrt{3}(r_+ - r_-)}
\end{eqnarray}
and the corresponding Hamiltonian is given as
\begin{equation}
H_E = \frac{\bar{n}e^{ - 3 r_0}}{24}\left[-(\bar{p}_0^2-\bar{p}_+^2-144m^2e^{4r_0})+\frac{12}{2\omega+3}p_\xi^2\right].
\end{equation}
A variation of action with respect to $n$ yields $\mathcal{H}_{E}=\frac{e^{  3 r_0}}{\bar{n}}H_E =0$. Again with similar canonical transformation, 
\begin{eqnarray}
\bar{T}=\frac{\xi}{p_\xi},\\
p_{\bar{T}}=\frac{p_\xi^2}{2},
\end{eqnarray}
Wheeler de Witt equation can be recast in following form: 
\begin{equation}
\left[\frac{\partial^2}{\partial r_0^2}-\frac{\partial^2}{\partial r_+^2}-144m^2e^{4r_0} -i\frac{24}{2\omega+3} \frac{\partial}{\partial \bar{T}}\right]\bar{\psi}=0.
\end{equation}

It is evident from the canonical transformation done in both the framework that $r_i=\beta_i$ and $T=\bar{T}$, hence the WDW equation in both the frame comes out to be exactly same and leads to similar wave packet if we transform  $\bar{a},\bar{b},\bar{c}$ back to the original $\sqrt{\phi} a,\sqrt{\phi} b,\sqrt{\phi} c$ respectively.\\

\subsection{Bianchi IX}
A Bianchi-IX metric given as
\begin{widetext}
\begin{equation}
ds^2 = n^2(t)dt^2 - a^2(t)dr^2-b^2(t)d \theta^2 -[a^2(t) \cos^2 \theta+b^2(t)\sin^2 \theta]d \phi^2.
\end{equation}
\end{widetext}
In the Einstein frame, the metric components are indicated with an overhead bar. \\

With the following change of variables
 \begin{eqnarray}
b(t)&=&e^{-\alpha} \frac{\beta}{a},\\
a(t)&=&e^{-\alpha/2 } a_0,\\
\phi(t) &=& e^{\alpha}
\end{eqnarray}
in the Jordan frame and following change 
\begin{equation}
 \bar{\beta}=\bar{a}\bar{b}
\end{equation}
in the Einstein frame coupled with a canonical transformation as done in the previous two cases, one can write down the WdW equations in the two frames. The result is the same, i.e., similar wave packet will be obtained if we revert the $\bar{a},\bar{b}$ to original $\sqrt{\phi} a, \sqrt{\phi} b$ respectively.

\section{Models with time dependent spatial curvature}

The equivalence of the two frames can also be shown in other Bianchi models as well. Bianchi I, V and IX are in fact examples of anisotropic models where the curvature of the space section is constant. But the conclusion remains the same for models with with time dependent spatial curvature as well. The example of a Bianchi type VI model can be taken up. Such a model is given by the metric 

\begin{equation}
ds^2 = n^2(t)dt^2 - a^2(t)dx^2-e^{- m x}b^2(t)dy^2 -e^{x}c^2(t)dz^2.
\end{equation}
With the change of variables as 
\begin{eqnarray}
a(t)&=&e^{-\alpha/2 +\beta_0 },\\
b(t)&=&e^{-\alpha/2 +\beta_0 +\sqrt{3}(\beta_+ -\beta_-)},\\
c(t)&=&e^{-\alpha/2 +\beta_0 -\sqrt{3}(\beta_+ -\beta_-)},\\
\phi(t)&=& e^{\alpha},
\end{eqnarray}

the Lagrangian and the Hamiltonian in the Jordan frame can be written respectively as
\begin{widetext}
\begin{equation}
L_j = \frac{e^{-\alpha / 2 + 3 \beta_0}}{n}\left[-6\dot{\beta}_0^2+6(\dot{\beta}_+  \\
-\dot{\beta}_-)^2-\frac{e^{-2\beta_0}n^2(m^2-m+1)}{2}+\frac{2\omega +3}{2}\dot{\alpha}^2 \right],
\end{equation}

and
 
\begin{equation}
H_j =  \frac{ne^{\alpha / 2 - 3 \beta_0}}{24}\left[-[p_0^2-p_+^2-12(m^2-m+1)e^{4\beta_0}]+\frac{12}{2\omega+3}p_\alpha^2\right].
\end{equation}

If we consider a canonical transformation $(\alpha,p_\alpha)$ to $(T,p_T)$ as
\begin{eqnarray}
T&=&\frac{\alpha}{p_\alpha},\\
p_T&=&\frac{p_\alpha^2}{2},
\end{eqnarray}

then with the Hamiltonian constraint, $\mathcal{H}=\frac{e^{-\alpha / 2 + 3 \beta}H}{n}=0$, WDW equation can be written as
\begin{equation}
\label{b6wdw-j}
\left[\frac{\partial^2}{\partial \beta_0^2}-\frac{\partial^2}{\partial \beta_+^2}-12(m^2-m+1)e^{4\beta_0} -i\frac{24}{2\omega+3} \frac{\partial}{\partial T}\right]\psi=0.
\end{equation}

\end{widetext}
In the Einstein frame, the metric is written as

\begin{equation}
ds^2 = \bar{n}^2(t)dt^2 - \bar{a}^2(t)dx^2-e^{-mx}\bar{b}^2(t)dy^2-e^{x}\bar{c}^2(t)dz^2,
\end{equation}

where the barred metric components are related to the unbarred components in the Jordan frame as ${\bar g}_{\mu\nu} = \phi g_{\mu\nu}$. With the transformation
\begin{eqnarray}
\bar{a}(t)&=&e^{r_0 },\\
\bar{b}(t)&=&e^{r_0 +\sqrt{3}(r_+ -r_-)},\\
\bar{c}(t)&=&e^{r_0 -\sqrt{3}(r_+ -r_-)},
\end{eqnarray}

The Lagrangian and the Hamiltonian in the Einstein frame look respectively as
\begin{widetext}
\begin{equation}
L_E=\frac{e^{3r_0}}{\bar{n}}\left[-6\dot{r}_0^2+6(\dot{r}_+-\dot{r}_-)^2-\frac{e^{-2r_0}n^2(m^2-m+1)}{2}+\frac{2\omega+3}{2}\dot{\xi}^2\right],
\end{equation}

and 

\begin{equation}
H_E = \frac{\bar{n}e^{ - 3 r_0}}{24}\left[-[\bar{p}_0^2-\bar{p}_+^2-12(m^2-m+1)e^{4r_0}]+\frac{12}{2\omega+3}p_\xi^2\right].
\end{equation}
\end{widetext}
Again with similar canonical transformation as
\begin{eqnarray}
\bar{T}&=&\frac{\xi}{p_\xi},\\
p_{\bar{T}}&=&\frac{p_\xi^2}{2},
\end{eqnarray}

and the Hamiltonian constraint  $\mathcal{H_E}=\frac{e^{  3 r_0}}{\bar{n}}H_E =0$, the WDW equation can be written as
\begin{equation}
\left[\frac{\partial^2}{\partial r_0^2}-\frac{\partial^2}{\partial r_+^2}-12(m^2-m+1)e^{4r_0} -i\frac{24}{2\omega+3} \frac{\partial}{\partial \bar{T}}\right]\bar{\psi}=0.
\end{equation}
One can easily show that $r_i=\beta_i$ and $T=\bar{T}$ from canonical transformation effected in both the framework, thus WDW equation in both the frame is exactly same, henceforth leads to similar the wave packet upon transforming $\bar{a},\bar{b},\bar{c}$ back to original $\sqrt{\phi} a,\sqrt{\phi} b,\sqrt{\phi} c$ respectively as usual.

\section{Generic Scenario with Concluding Remarks}

By a mathematical equivalence, we mean any physical quantity obtained in Jordan frame can be mathematically transformed into Einstein frame and yields the same value as we would have got, should we do the calculation Einstein frame to start with, and vice-versa.\\

Quantum mechanically, this mathematical equivalence seems to be broken as reported in literature in several times\cite{cho, nb}. Here we claim that this reported in-equivalence has to do with inconsistent operator ordering. Given a frame, be it Einstein or Jordan, we always have to choose a particular ordering of operators to go over to quantum theory. Now, choosing one particular operator ordering in Jordan frame fixes the operator ordering in Einstein frame and vice versa. This strips us off the freedom of choosing operator ordering in both frames independently. Should we do operator ordering in both frames arbitrarily, it might become inconsistent with each other, leading to different quantum Hamiltonian and hence distinct behavior of the wave packets. Thus, the apparent discripency should not be attributed to quantum effects, rather they are artifact of inconsistent operator ordering. What we do in this work is basically picking up a consistent choice of operator ordering in both frames. We do the parametrization in a clever way, which renders the most natural choice of operator ordering to be the consistent ones. \\

It deserves mention that parametrization in this particular way and the natural operator ordering in new variables $\beta_{0},\beta_{\pm}$ indicate a particular choice of operator ordering in terms of old variables $a,b,c$. This choice is by no means a unique one and it is not required to prove the equivalence either.
For homogeneous models, however, there is a natural unique  choice of the parametrization as shown in what follows.\\

 In Jordan frame, we parametrize the metric in following way: 
\begin{equation}
g_{ij}= e^{-\alpha} h_{ij}, \\
\phi=e^{\alpha}.
\end{equation}

Now, since we are dealing with homogeneous models, we parametrize $h_{ij}$ by some function of time in some particular way. It is easy to see 
\begin{equation}
\mathcal{L}_{J}=ne^{-\frac{\alpha}{2}} \sqrt{h} \left[R_{h_{ij}} + \frac{2\omega +3}{2}\dot{\alpha}^{2}\right]
\end{equation}

From this we write down the Hamiltonian and subsequent Hamiltonian constraint and quantize the theory in canonical manner.\\

Now, the Einstein frame is a conformal transformation of Jordan frame. But homogeneity guarantees that $\phi$ is a function of $t$ alone, hence the conformal factor is a function of $t$ alone. This implies going from Jordan to Einstein frame does not change the Bianchi class of the model concerned. Had it been the case that $\phi$ is a function of space as well as time, things would have been more intricate and this method would have broken down i.e the proof would not have gone through.\\

So, in Einstein frame, we parametrize the metric $\bar{g}_{ij}$ in same way as we have parametrized the metric $h_{ij}$. This will lead to
\begin{equation}
\mathcal{L}_{E}=\bar{n}\sqrt{-\bar{g}}\left[\bar{R}+\frac{2\omega +3}{2}\dot{\alpha}^{2}\right]
\end{equation}

Since, $\bar{g}_{ij}=\phi g_{ij}=h_{ij}$, we have $\bar{R}=R_{h_{ij}}$, and parametrization in similar way guarantees that consistent operator ordering is being chosen automatically, hence the equivalence becomes a obvious one. \\

It deserves mention that this proof is really independent of how we are defining the time variable for quantum theory, only a properly oriented time parameter is required.  Nonetheless, one might wonder that the choice of parametrization or the transformation, given by Eq.~\eqref{conformal} to make the Lagrangian diagonal is actually effecting a conformal transformation to Einstein frame in disguise. In what follows, we will refer to the section where we have discussed Bianchi-I, but this is generically true for all homogenous models.\\

It is a striaght forward exercise to extend the prescription for a cosmological model with fluid. It deserves mention that the application of {\it S-theorem} \cite{sb} and resulting anomalous symmetry breaking in quantum FRW cosmological model with radiation matter content \cite{sp6} is technically done using the same transformation. Hence, it is worthwhile to check whether such symmetry breaking can happen in Jordan frame as well before effecting the canonical transformation and confirm the quantum equivalence. \\
 
It also deserves mention that we have tried other transformations involving the Brans-Dicke parameter $\omega$, and at least for some particular vaules of $\omega$ could write down separable expressions. One  difficulty is that of picking up a nice time variable, whose corresponding momentum appears only in the first order. But the most important problem is that it is impossible to pick the corresponding operator ordering in the two frames, that is irrecoverably lost in the transformation.\\

We hope the work will clarify the issue regarding equivalence prevailing in literature. The question of mathematical equivalence, thus being clarified, one could ask for more sophisticated questions like quantum mechanically which frame is more physical and how the quantum behavior changes once we go from one frame to another frame. There is some work in this connection\cite{cho}, but we hope to get back to these questions in future with further informations.

\section{Acknowledgement}
This work of S. Pal was supported by the US Department of Energy under contract DE- SC0009919. S. Pandey was supported by CSIR, India. The authors thank Abhik Sanyal and Nayem Sk for pointing out certain problems in the transformation. The authors thank Ott Vilson for useful comments on the manuscript and providing convincing argument that there are no problems indeed in a private correspondence.

\end{document}